\documentclass{eas}
\usepackage{graphicx}
\begin{document}

\title{Quenching factor measurement in low pressure gas detector for directional dark matter search} 
\author{O. Guillaudin}\address{Laboratoire de Physique Subatomique et de Cosmologie,
		Universit\'e Joseph Fourier Grenoble 1,
		CNRS/IN2P3, Institut Polytechnique de Grenoble,
		53, rue des Martyrs, Grenoble, France}
\author{J. Billard}\sameaddress{1}
\author{G. Bosson}\sameaddress{1}
\author{O. Bourrion}\sameaddress{1}
\author{T. Lamy}\sameaddress{1}
\author{F.~Mayet}\sameaddress{1} 
\author{D. Santos}\sameaddress{1}
\author{P. Sortais}\sameaddress{1}

\begin{abstract}
There is considerable experimental effort dedicated to the directional detection of particle dark matter. 
Gaseous $\mu$TPC detectors present the privileged features of being able to reconstruct the track and the energy 
of the recoil nucleus following the interaction. A precise measurement of the recoil energy is a key point for the directional search strategy. Quenching has to be taken into account, i.e.
only a certain fraction of the recoil energy is deposited in the ionization channel. Measurements of the ionization quenching factor for different gas mixture at room temperature have been made with a dedicated ion beam facility at the LPSC
of Grenoble.
\end{abstract}
\maketitle
\section{Introduction}
The MIMAC Collaboration (Santos {\em et al.\/} \cite{mimac}) aims at building a directional Dark Matter detector composed of a matrix of pixelised Micromegas Detectors. Directional detection requires the measurement of both 3D track and energy of recoiling nuclei (Ahlen {\em et al.\/} \cite{Ahlen}; Billard {\em et al.\/} \cite{Billard}). The recoil energy measurement is in this case a challenging issue as the scattering of a Dark Matter particle produces, in most cases, a few keV recoiling nucleus. The energy released by a particle in a medium produces in an interrelated way three different processes:
\begin{enumerate}
  \item ionization, producing a number of electron - ion pairs,
  \item scintillation, producing a number of photons through de-excitation of quasimolecular states
  \item heat produced essentially by the motion of nuclei and electrons.
\end{enumerate}

The fraction of energy given to electrons has been estimated theoretically by (Linhard {\em et al.\/} \cite{Lind}) and may be simulated  by the SRIM Monte Carlo code (J.~Ziegler {\em et al.\/} \cite{Ziegler}).
The ionization quenching factor (IQF) is defined as the fraction of energy released through
ionization by a recoil in a medium compared with its kinetic energy. Measuring
IQF, especially at low energies, is a key point for Dark Matter detectors, since it
is needed to evaluate the nucleus recoil energy and hence the WIMP kinematics.

In the last decades an important effort has been made to measure the IQF in different
materials: gases (Verbinski \& Giovannini \cite{Verbi}), solids (Jones \& Kraner \cite{Jone},
Gerbier {\em et al.\/} \cite{Gerb}) and liquids (Aprile {\em et al.\/} \cite{Apri}), using different techniques.
The use of a mono energetic neutron beam has been explored in solids with success
(Jageman {\em et al.\/} 2005, Simon {\em et al.\/} 2003). However in the low energy range the
measurements are rare or absent for many targets (e.g. Helium or $\rm CF_{4}$) due to ionization
threshold of detectors and experimental constraints.

Our motivation for this work is to obtain the best reconstruction of the detected nuclear recoil energy in low 
pressure gaseous dark matter search experiments. 

\section{Experimental method and setup}
The IQF can be determined by measuring the ionization produced by a nuclear recoil of a known energy in the target material. We have developed an experimental setup devoted to the measurement of low energy
(keV) ionization quenching factor. The experimental set-up is the following: an Electron Cyclotron Resonance
Ion Source (ECRI) (Geller \cite{Gell}) with an extraction potential from a fraction of kV
up to 50 kV, is coupled to a Micromegas detector via a 1 $\mu$m hole with a differential pumping. This setup allows measurements from few mbar to 1 bar.

\subsection{The ion source}

\begin{figure}
\begin{center}
\includegraphics[width=12cm]{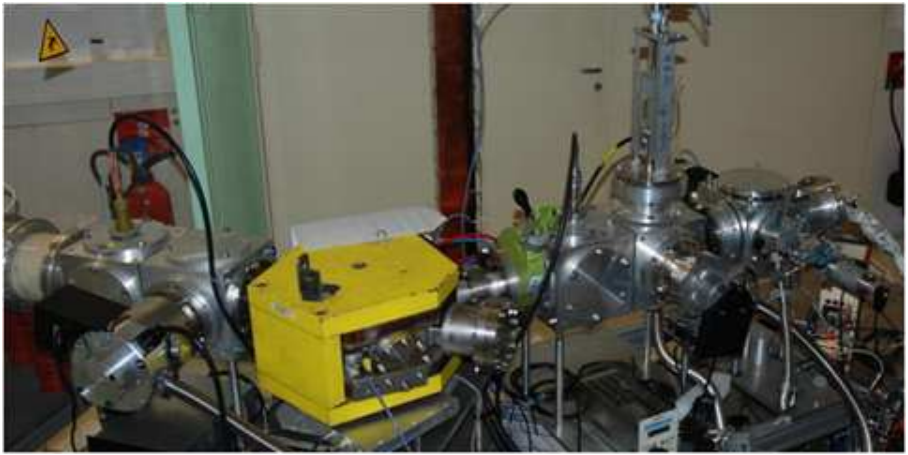}
\caption{ECRI ion source with 45 degrees spectrometer (yellow)}
\label{ECRI}
\end{center}
\end{figure}

\begin{figure}
\begin{center}
\includegraphics[scale=0.6]{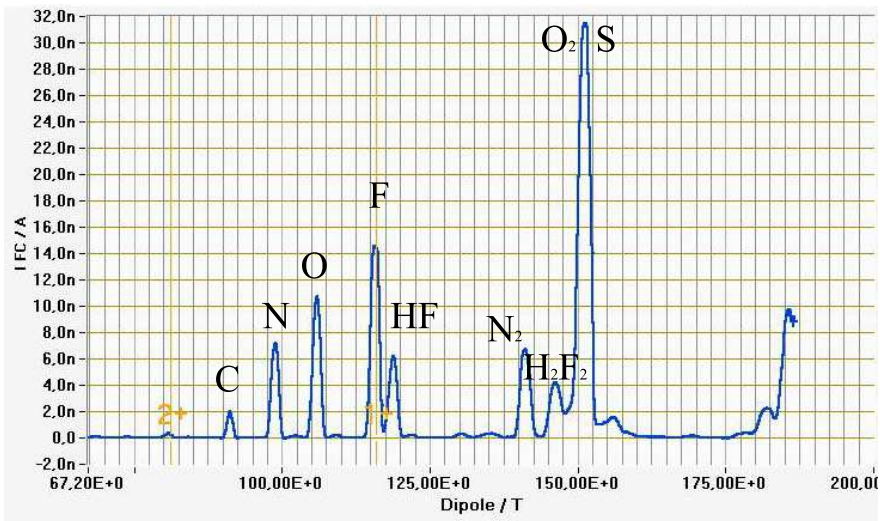}
\caption{Typical ion source spectrum using $\rm SF_{6}$ gas in the plasma chamber with 20 kV as extraction potential}
\label{SpectrSource}
\end{center}
\end{figure}

The ECRI ion source (Electron Cyclotron Resonace Source) (Fig. \ref{ECRI}) is a plasma device designed and built by the LPSC ion source team to provide charged ions at low velocities from a fraction of keV to 50 keV. One of the main features
of this facility is the availability of very low and stable currents of a few pA. In the current 
configuration, the beam is adapted to the 45 degrees spectrometer by an Einzel lens. A micrometer aperture (1 $\mu m$ in diameter) selects a part of this beam which is injected in the Micromegas detector and allows a rate of about 25 ions per second. Depending to the gas injected in the plasma chamber, the source is able to produce many type
of beams : $\rm ^{1}H$, $\rm ^{3}He$, $\rm ^{4}He$, $\rm^{12}C$,$\rm^{19}F$ and to separate the desired ion species (Fig. \ref{SpectrSource}).

The ions energy have been previously checked by time-of-flight measurements. A 50 nm thick Silicon Nitride 
membrane ($\rm Si_{3}N_{4}$) was used as interface to produce low energy electrons that has been detected by a first channeltron. A second channeltron was used to detect ions for six different known positions. This setup allowed to measure the energies of the ions just after the foil. With this configuration (Silicon Nitride membrane interface), a first energy measurement with Micromegas was performed. Consequently, we could verify that the energy measured by TOF were consistent with the extraction potential values in kV for the 1+ state of charge of ions with the 1 $\mu m$ diameter aperture interface.

\subsection{Detector calibration}
\begin{figure}
\begin{center}
\includegraphics[scale=0.3]{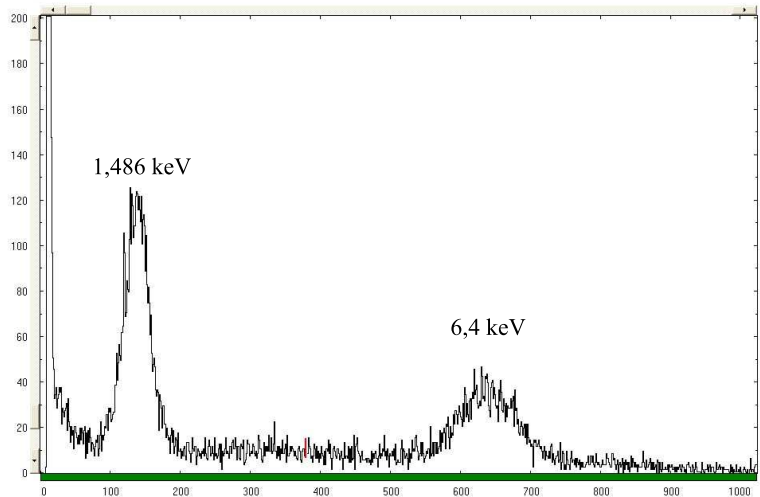}
\caption{Response of Micromegas filled with $\rm ^{4}He + 5\% \ C_{4}H_{10}$  at 1000 mbar to fluorescence X-rays.  
Left peak: Aluminium X-rays  at 1.486 keV with measured resolution of 29.4 \% FWHM
Right peak: Iron X-rays  at 6.4 keV with measured resolution of 15.3 \% FWHM}
\end{center}
\end{figure}

The Micromegas used in this work was a standard Bulk Micromegas (Giomataris {\em et al.\/} \cite{Giom}), in which 
the mesh and the anode are built and integrated with a fixed 192 $\mu m$ gap. The active area is $\rm 100 \times 100 \, mm^{2}$. The gap between the copper anode and stainless steel mesh is maintained by cylindrical pillars every 2~mm. This gap is well adapted for gas pressure between 350 and 1300 mbar with Helium and 50 mbar with $\rm CF_{4}$. The drift distance between the cathode and the mesh is 4 cm, which is large enough to include the tracks of nuclei of energies up to 50 keV at 350 mbar. These tracks, of the order of 6 mm for 50 keV, are roughly of the same length than the electrons tracks 
produced by a 6 keV  X-rays. We used an individual power supply for each electrode (mesh and anode) with cathode drift to the ground, allowing to independently vary the different fields. For coupling reasons with the ion source, the $\rm \mu m$ hole interface (cathode) must be set to the ground.

The detector is exposed to different radioactive sources producing X-rays from 1.486 keV to 6.4 keV and the measurements were performed  in different gas mixtures. Fluorescence targets combined with an Alpha source are used to produce 1.486 keV (Aluminium) or 6.4 keV (Iron). A standard $\rm ^{55}Fe$ X-ray source providing the 5.9 keV X-Rays has also been used for calibration. 
The signals from the Micromegas anode were processed by the analogue readout composed of a dedicated charge sensitive pre-amplifier and of a commercial spectroscopy amplifier. Due to the low drift velocity of electron in such a low drift field
(80 V/cm in Helium mixture), the decay time constant of this pre-amplifier is large enough (few hundred $\rm \mu s$) to ensure a total collection charge and an output proportional to the total energy of the event. The Spectroscopy amplifier output is connected to a FAST ComTec MCA-3 PCI-bus Multichannel Analysers (MCA) for spectra building. Each spectrum was fitted to get the peak positions and the energy resolution. The linearity was checked using different X-rays energies between 1.5 and 8 keV.

\begin{figure}
\begin{center}
\includegraphics[scale=0.5]{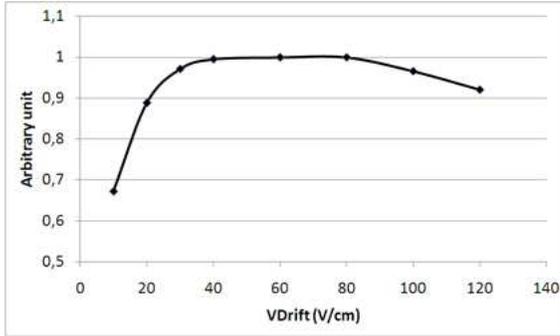}
\caption{Dependence of the electron transmission with the drift field for the readout with a 192 $\mu m$-thick amplification gap for $\rm ^{4}He + 5\% \ C_{4}H_{10}$ gas mixture at 350 mbar }
\label{Transparence}
\end{center}
\end{figure}

In order to obtain the maximum electron collection efficiency, the drift voltage was varied for a fixed mesh voltage.
The maximum electron collection efficiency is reached for an optimal ratio between amplification and drift electric field. Fig.~\ref{Transparence} shows the typical plateau for Micromegas readout planes where the maximum electron transmission is obtained for a field value of about 60 V/cm in Helium mixture at 350 mbar and 50 V/cm in pure $\rm CF_{4}$ at 50 mbar. 

\begin{figure}
\begin{center}
\includegraphics[scale=0.4]{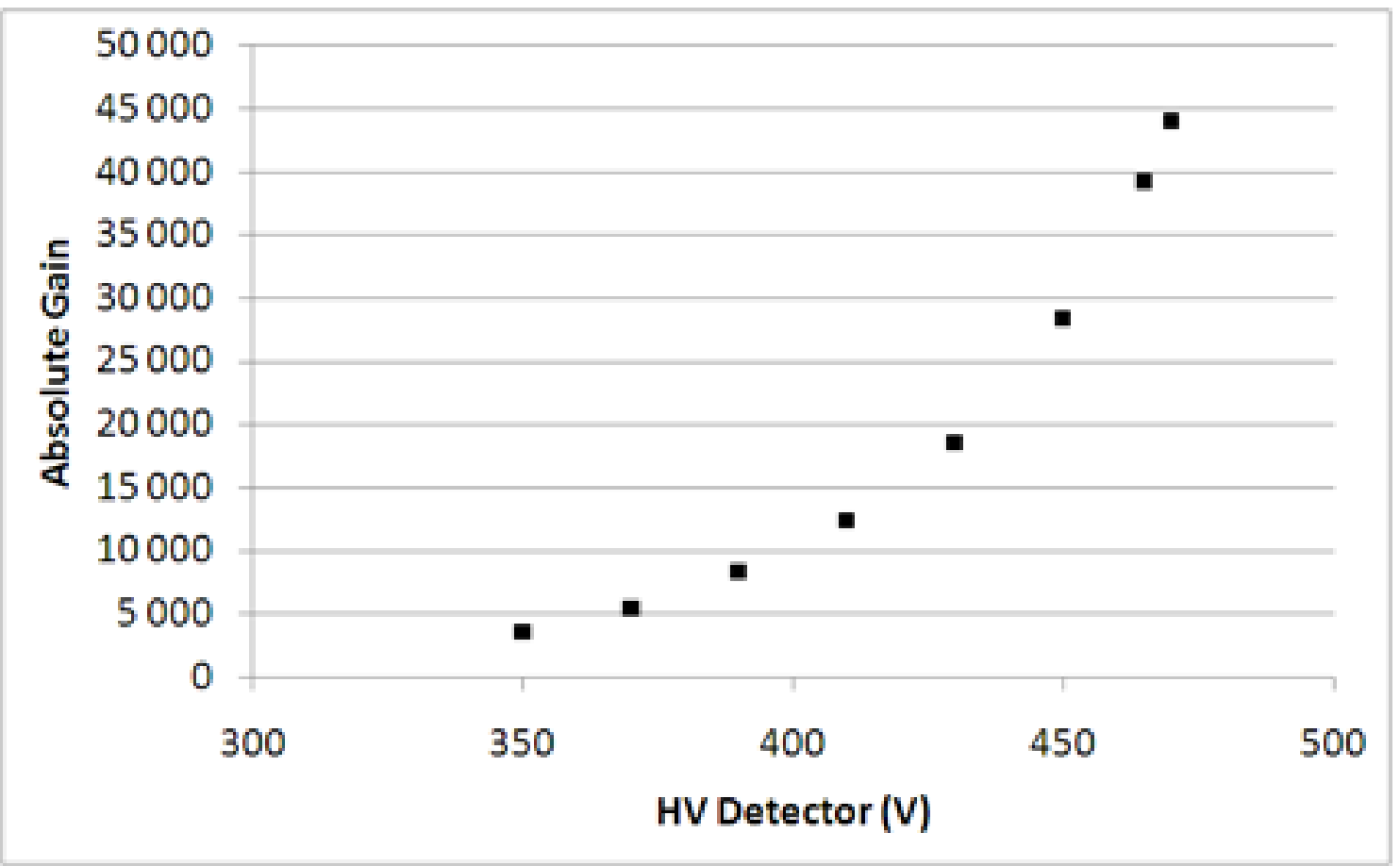}
\includegraphics[scale=0.4]{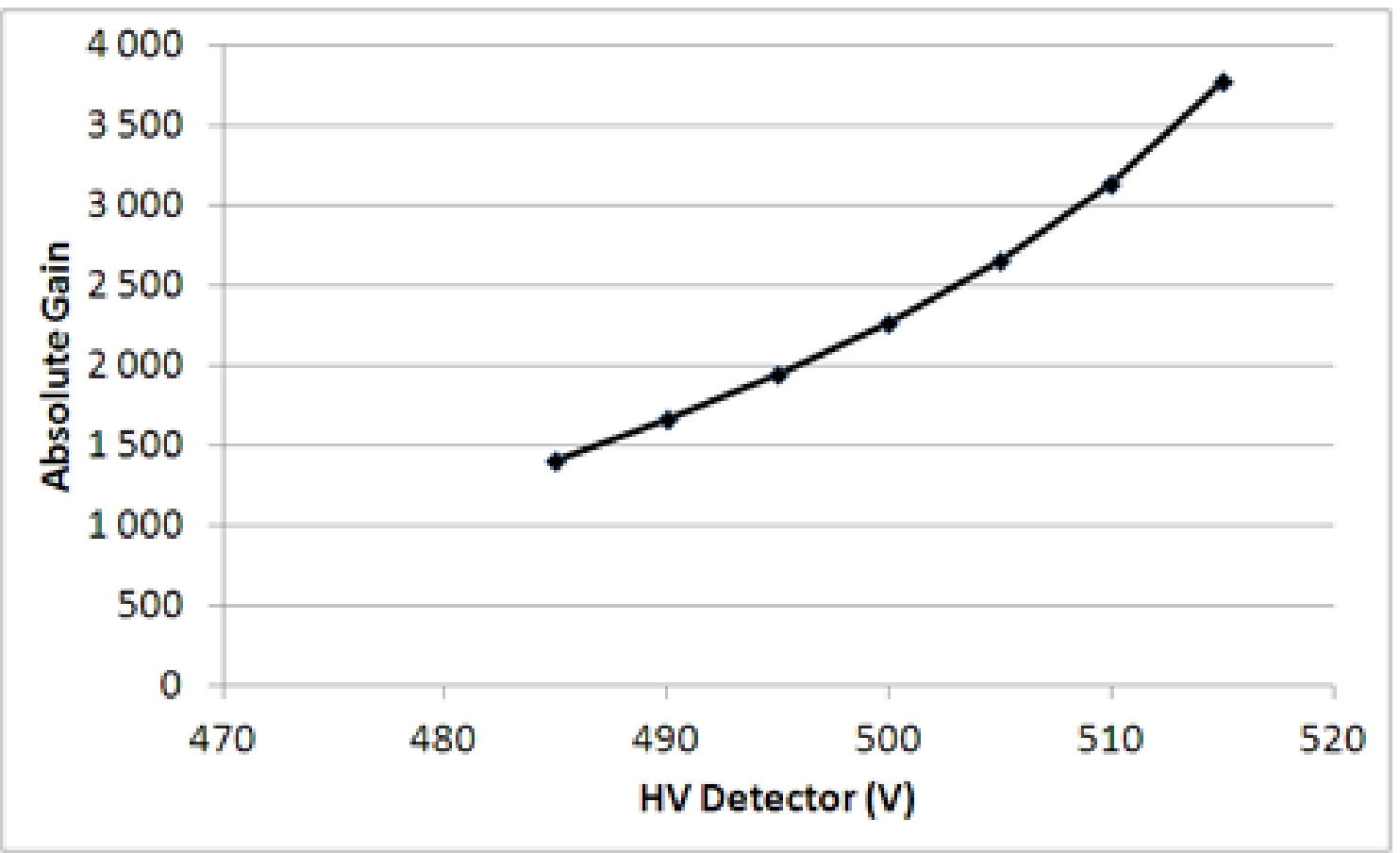}
\caption{Absolute gain as a function of the High Voltage detector for 192 $\mu m$-thick amplification gap obtained with a iron $\rm ^{55}Fe$. Left, with  $\rm ^{4}He + 5\% \ C_{4}H_{10}$ gas mixture at 350 mbar and right, with pure $\rm CF_{4}$ gas at 50 mbar}
\label{Gain}
\end{center}
\end{figure}

After this, the drift field was fixed in the region where the detector presents the maximum electron transmission and 
the mesh voltage was varied to obtain the gain curves shown in fig. \ref{Gain}. A maximum gas gain greater than $\rm 4.5\times10^{4}$  has been achieved in $\rm ^{4}He + 5\% \ C_{4}H_{10}$ mixture at 350 mbar and  $\rm 4\times10^{3}$ in pure $\rm CF_{4}$ at 50 mbar. In the same conditions, the magnitude of the energy resolution is about 16\% for Helium and 25\% for pure $\rm CF_{4}$. Those gain values are large enough to trigger the MIMAC front end electronic used for 3D track reconstruction.

\section{Ionization quenching factor measurements}
A precise methodology was used to measure the quenching factor of a given nucleus moving in the gas mixture. Before the measurement, the Micromegas detector was first exposed to X-rays sources to establish an electron equivalent energy signal. Then for the actual measurement, the number of ions injected per second was kept lower than 25, to prevent any problem of recombination in primary charge collection or space charge effect. The recoil IQF is the ratio between this measured energy and the kinetic energy of such nucleus.

\subsection{He Quenching factor}

\begin{figure}
\begin{center}
\includegraphics[scale=0.4]{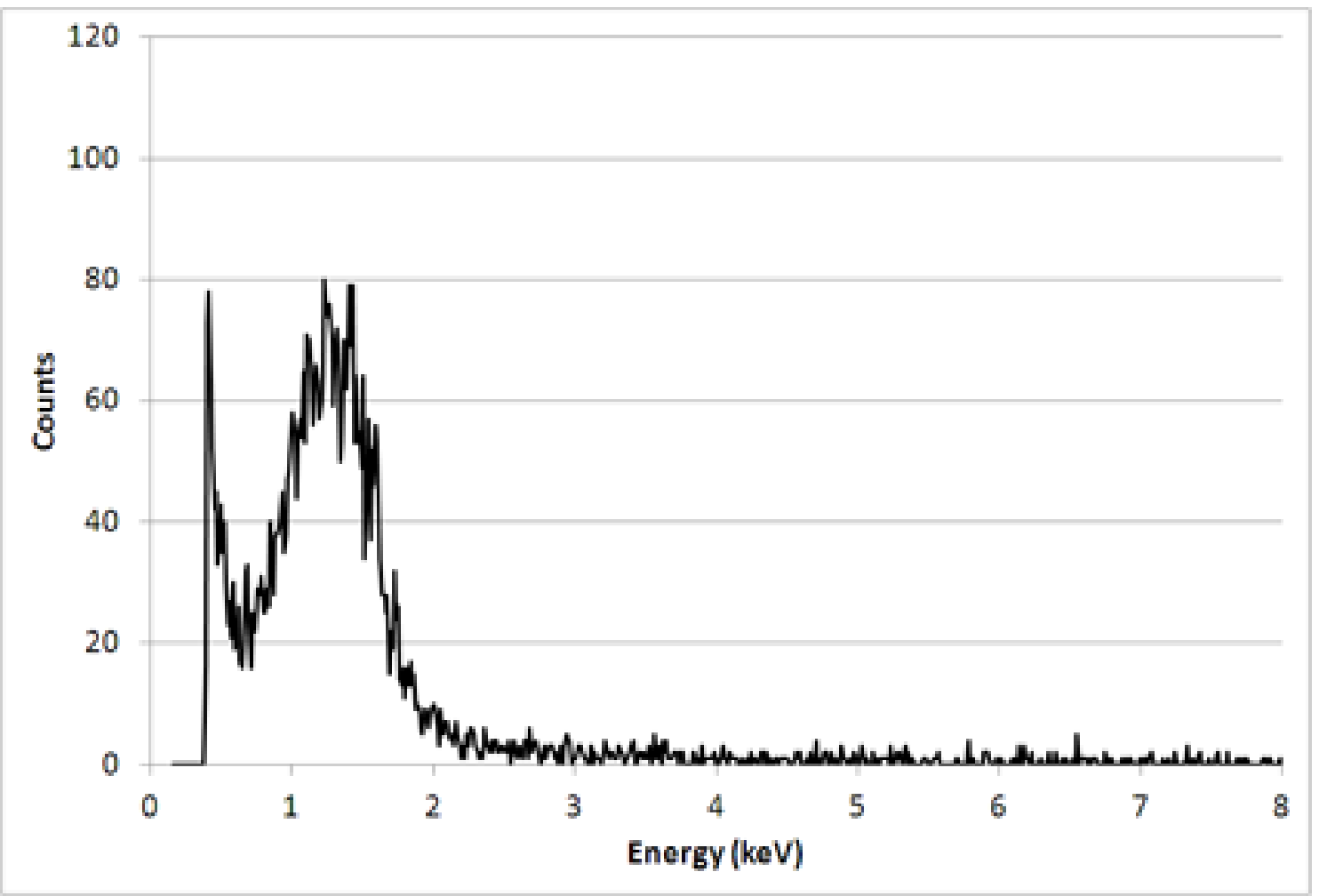}
\includegraphics[scale=0.4]{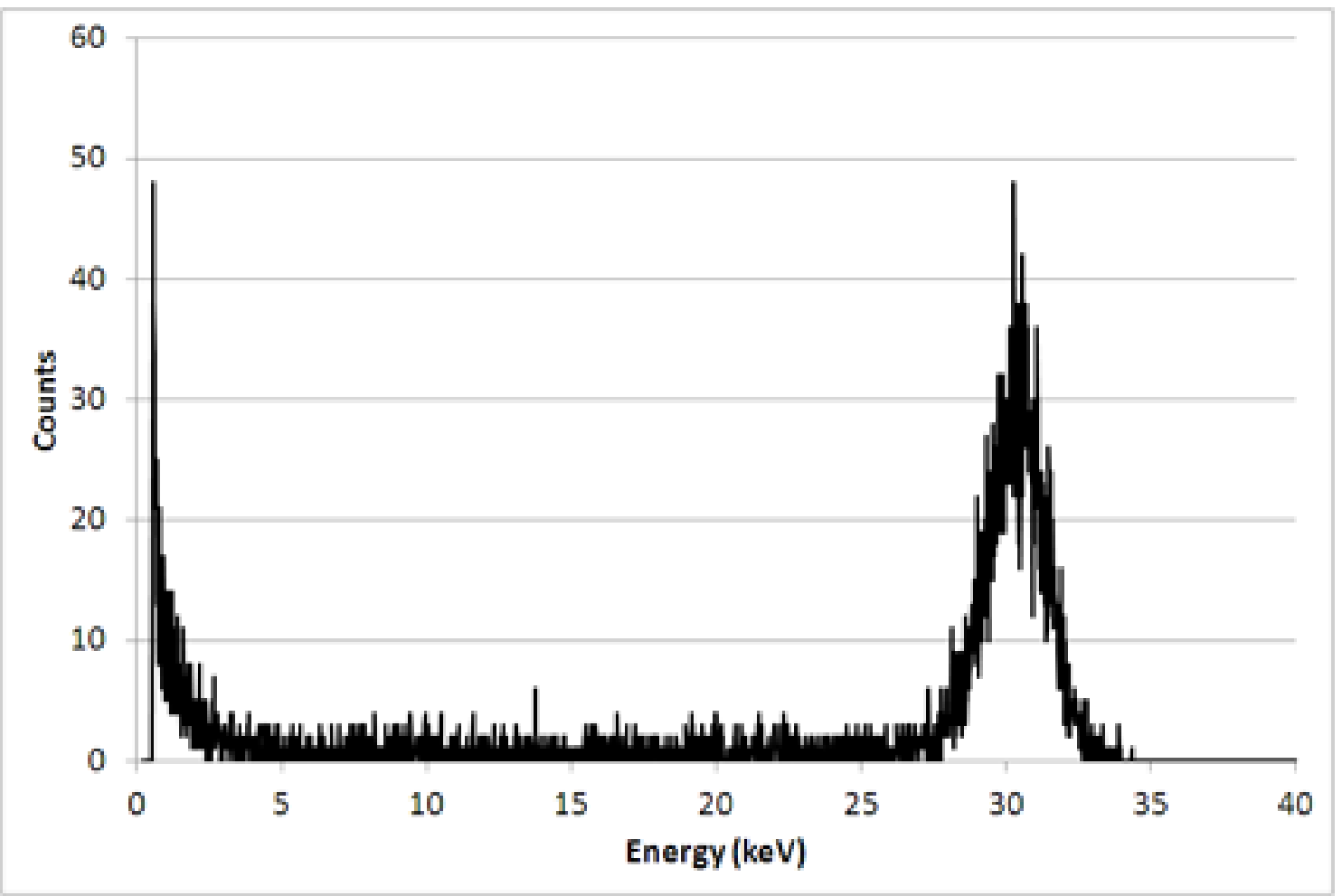}
\caption{Spectra of 3 keV  (left) and 40 keV  (right) kinetic energy $\rm ^{4}He$ in $^{4}He + 5\% \ C_{4}H_{10}$ mixture at 350 mbar.}
\label{HeQ3_40}
\end{center}
\end{figure}

The ionization spectra of 3 keV and 40 keV $\rm ^{4}He$ nuclei in $\rm He + 5\% \ C_{4}H_{10}$ gas mixture at 350 mbar are shown in Fig. \ref{HeQ3_40}. The measured ionization energy for 3 keV was 1.275 keV with an energy resolution of 54 \% (FWHM). At 40 keV, the measured ionization energy was 30.3 keV with an energy resolution of 7.6 \% (FWHM). The measurements have been focused on the low energy $\rm ^{4}He$ IQF. 

\begin{figure}[t]
\begin{center}
\includegraphics[scale=0.4,angle=270]{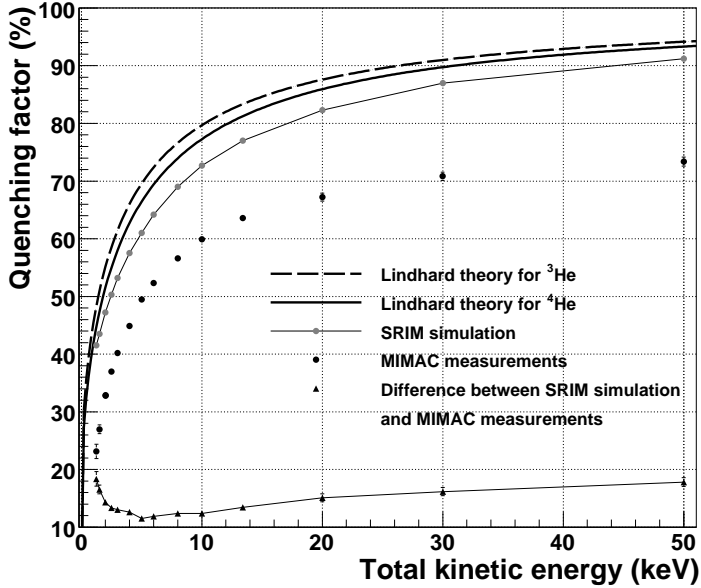}
\caption{Comparison between Lindhard theory, SRIM simulation and Quenching measurements in  $\rm ^{4}He + 5\% \ C_{4}H_{10}$ gas mixture at 700 mbar (from Santos {\em et al.\/} \cite{Sant})}
\label{HeQ}
\end{center}
\end{figure}

Fig. \ref{HeQ} presents the results for $\rm ^{4}He$ in the  $\rm He + 5\% \ C_{4}H_{10}$ gas mixture at a pressure of 700 mbar for the energy range 1.5 and 50 keV. A Lindhard theory for $\rm ^{4}He$ ions in pure $\rm ^{4}He$ and SRIM simulation are also plotted. 
A difference between the SRIM simulation and the experimental points of
up to 20\% of the kinetic energy of the nuclei can be observed. This difference may be assigned to the scintillation produced by the $\rm ^{4}He$ nuclei in $\rm ^{4}He$ gas.

As shown in (Trichet {\em et al.\/} \cite{Trich}), the energy resolution of Micromegas μTPC has been measured down to 1 keV. Threshold as low as 300 eV (ionization) has been reached.
Measurements have been done in various experimental conditions, showing a clear increase of the IQF at lower gas pressure  (Mayet {\em et al.\/} \cite{Mayet}).

\begin{figure}
\begin{center}
\includegraphics[scale=0.5]{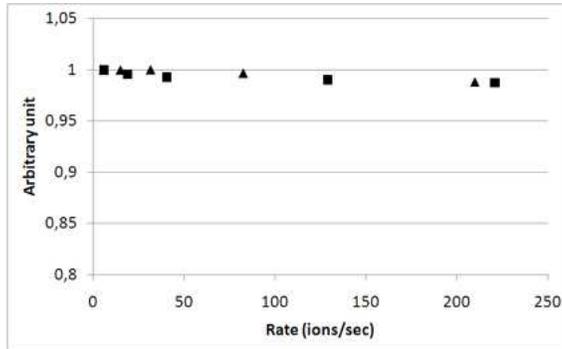}
\caption{Ionization of 20 keV $\rm ^{4}He$ recoils as a function of count rate in  $\rm ^{4}He + 5\% \ C_{4}H_{10}$ gas mixture at 350 mbar. Square for ion beam perpendicular to Micromegas plane, triangle for ion beam at $\rm 30^\circ$ to the Micromegas plane}
\label{He_rate}
\end{center}
\end{figure}

To be sure to exclude a problem of recombination in primary charge collection or space charge effect in Micromegas amplification region, some measurements with different ion rates have been performed at various recoil energies. The observed variation is less than 2 \% between 1 and 200 ions/s and should be considered as constant as shown in Fig. \ref{He_rate} for 20 keV ion energy and for two beam injection configurations: perpendicular ion beam and angle of  $30^\circ$ relative to the Micromegas plane.

\subsection{CF4 Quenching factor}
\begin{figure}
\begin{center}
\includegraphics[scale=0.4]{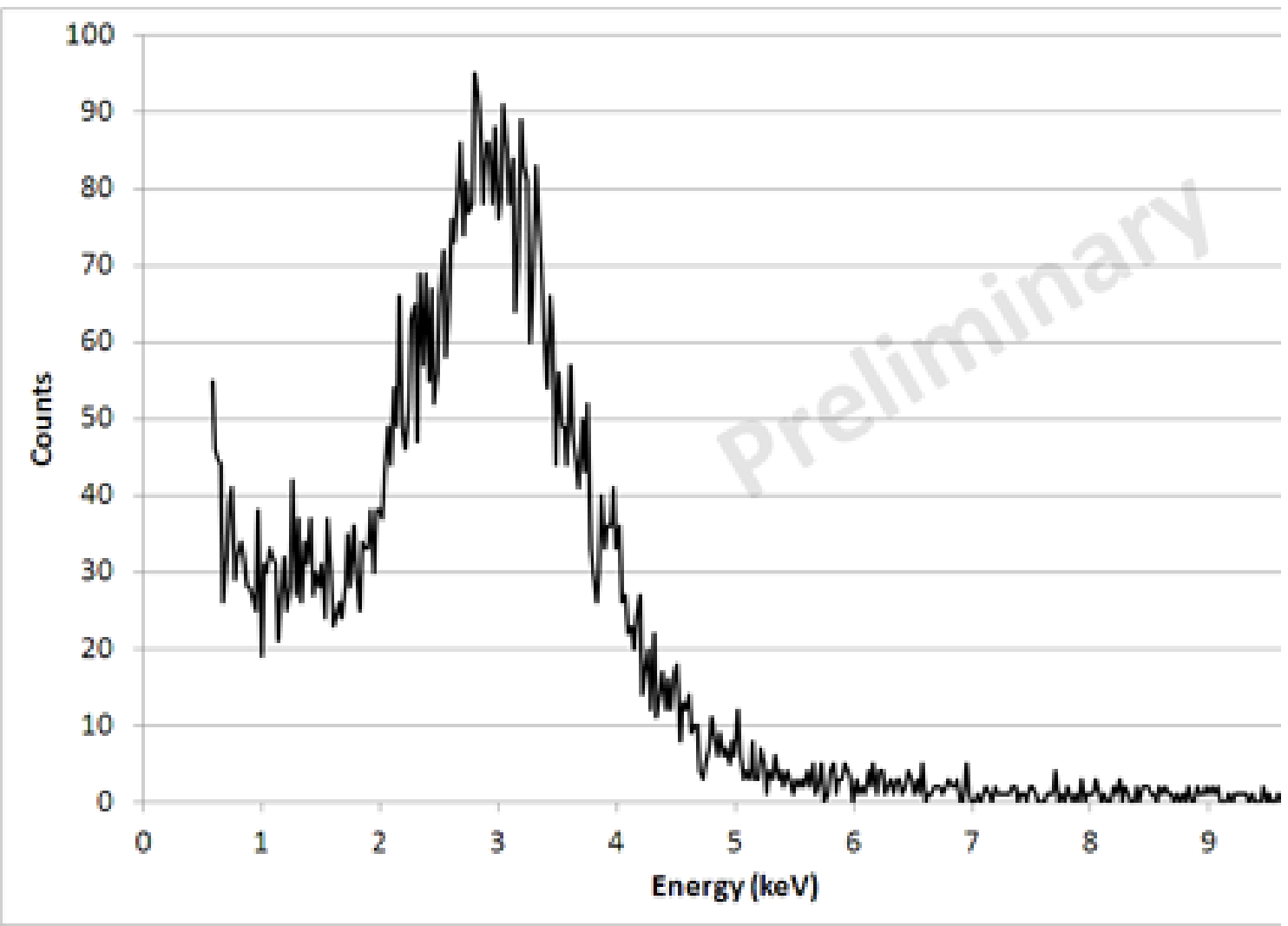}
\includegraphics[scale=0.4]{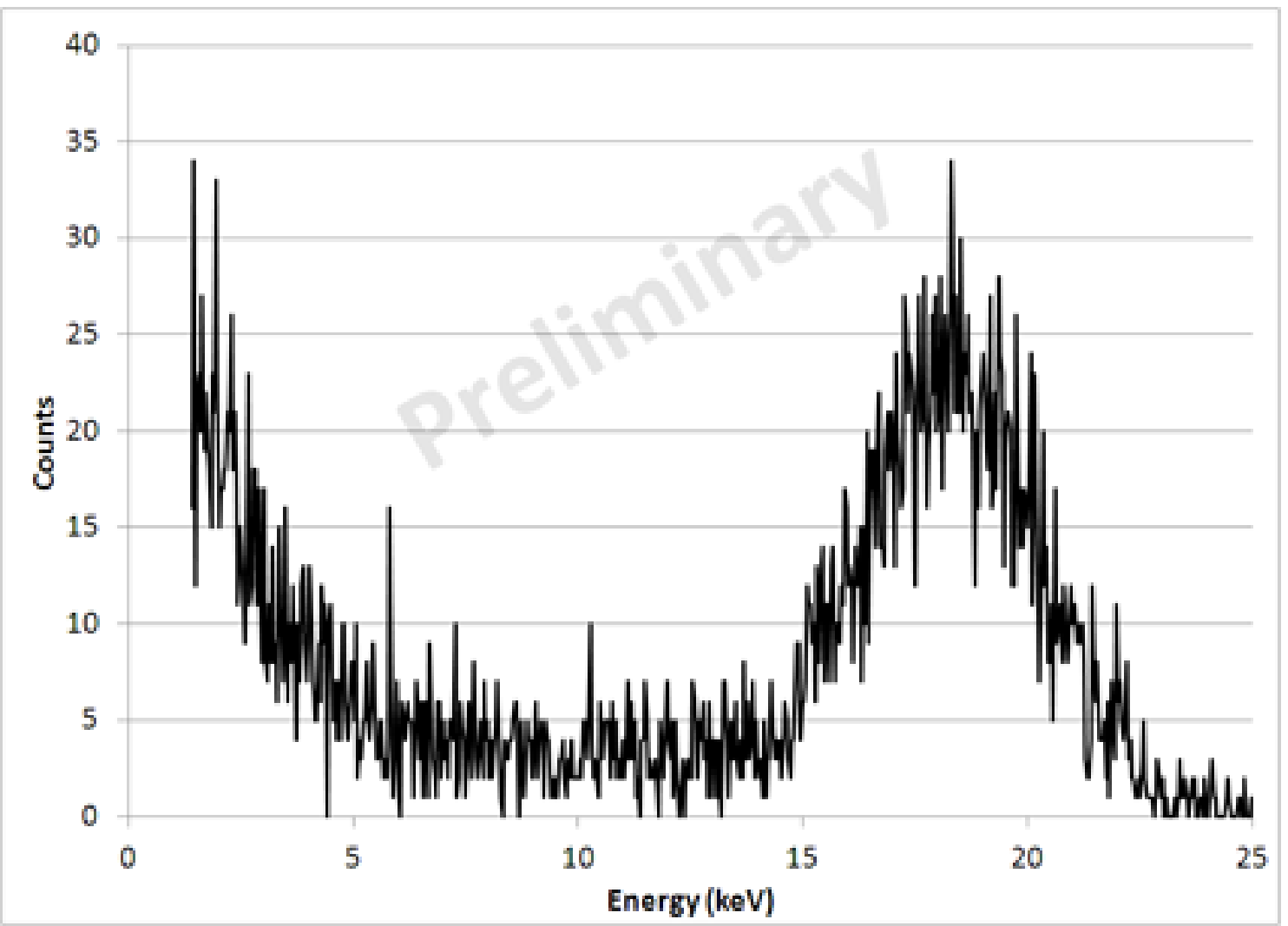}
\caption{Preliminary spectra of 10 keV  (left) and 40 keV  (right) kinetic energy $\rm ^{19}F$ in pure $\rm CF_{4}$ at 50 mbar.}
\label{F_S10_40}
\end{center}
\end{figure}

\begin{figure}
\begin{center}
\includegraphics[scale=0.4]{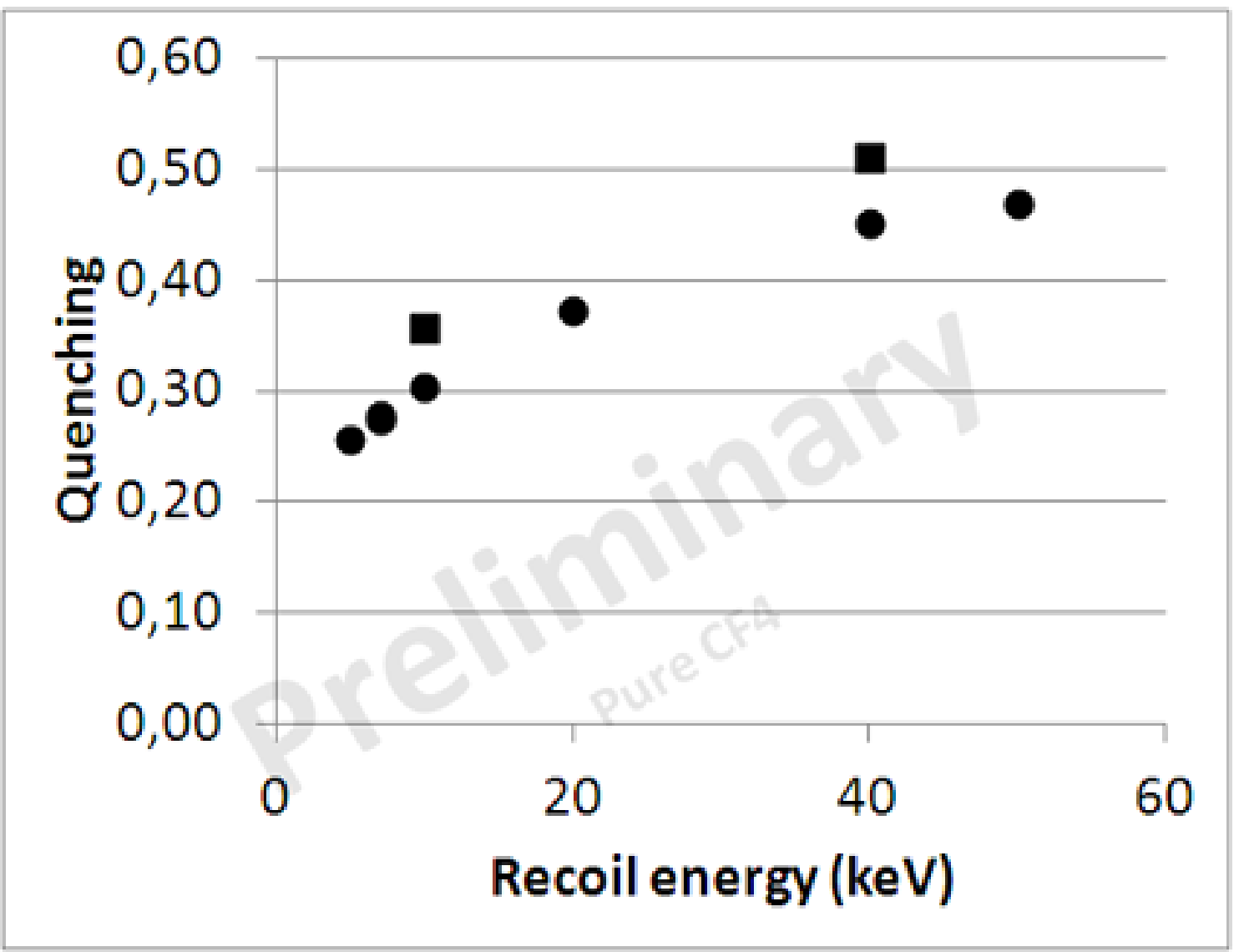}
\includegraphics[scale=0.4]{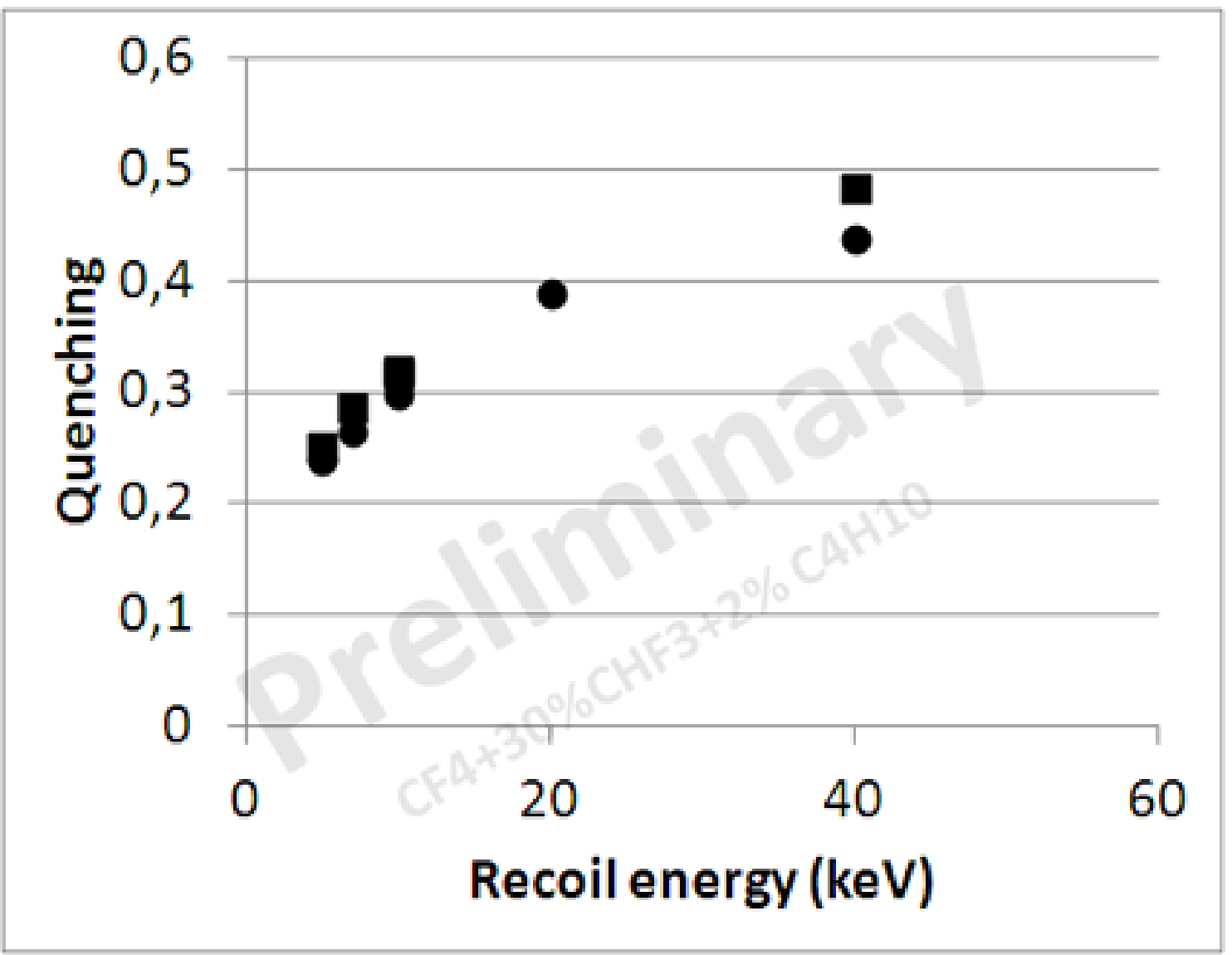}
\caption{Comparison between Fluorine (circle) and Carbon (square) Quenching factor in pure $\rm CF_{4}$ at 50~mbar (left) and in ($\rm CF_{4} + 30\% CHF_{3} + 2\% C_{4}H_{10}$) gas mixture at 50~mbar (right).}
\label{F_Quench_1_40}
\end{center}
\end{figure}

The ionization spectra of 10 keV and 40 keV $\rm ^{19}F$ nuclei in pure $\rm CF_{4}$ gas are shown in Fig. \ref{F_S10_40}. With nuclei at 10 keV, the ionization energy was measured at 2.9 keV with energy resolution of 49\% (FWHM). At 40 keV, the measured ionization energy was 18.2 keV with energy resolution of 24\% (FWHM). The measurements have been focused on the low energy $\rm ^{19}F$ and $\rm ^{12}C$ IQF in pure $\rm CF_{4}$ and $\rm CF_{4}$ gas mixture ($\rm CF_{4} + 30\% CHF_{3} + 2\% C_{4}H_{10}$) at 50 mbar as shown in Fig. \ref{F_Quench_1_40} . Quenching factor is still comfortable even at low energy, and few keV recoil energy could be measured with Micromegas at low pressure. 

Adding $\rm CHF_{3}$ in $\rm CF_{4}$ does not change much the quenching factor but reduce electron drift velocity. However, considering that the quenching of Fluorine and Carbon are very close, it would be very difficult to separate $\rm ^{12}C$ and $\rm ^{19}F$ ions with a track/energy analysis.

\section{Summary}
In summary, the possibility to measure low energy $^{4}He$ and $^{19}F$ recoils with Micromegas detectors at low pressure has been demonstrated. First measurements of Fluorine recoils in $\rm CF_{4}$ gas mixture have been presented and the results confirm the possibility to develop $\mu$TPC with precise energy measurement using Fluorine as target material for Dark Matter search. The addition of $CHF_{3}$ used for slowing down the electron drift velocity does not change the gain and Quenching Factor. Micromegas still shows good performances even at low pressure. In any case the results obtained up to now validate the MIMAC concept for the construction of a large TPC for directional detection of Dark Matter. It was recently shown that using Micromegas with a greater amplification space ($\rm 256\,\mu m$ instead of 
$\rm 192\,\mu m$) the gain and energy resolution could be improved. This work is still under investigation.


\end{document}